\documentclass[aps,prl,superscriptaddress,twocolumn,showpacs]{revtex4}
\usepackage{amsmath,epsfig,color,amssymb}
\usepackage{graphicx}
\usepackage{graphicx}
\usepackage{dcolumn}
\usepackage{bm}
\usepackage{graphicx}
\usepackage{amsmath, amsfonts, amssymb,mathrsfs}
\usepackage{pstricks}
\usepackage{amsxtra}
\usepackage{amsthm}
\usepackage{natbib}

\def\be{\begin{equation}}
\def\ee{\end{equation}}
\def\bea{\begin{eqnarray}}
\def\eea{\end{eqnarray}}

\def\be{\begin{equation}}      
\def\ee{\end{equation}}
\def\beu{\begin{equation*}}   
\def\eeu{\end{equation*}}

\providecommand{\del}{\partial}

\graphicspath{{graphQNLfigures/}}

\begin{document}
\title{Single-photon nonlinear optics with graphene plasmons}

\author{M.~Gullans}
\affiliation{Department of Physics, Harvard University, Cambridge, MA 02138, USA}
\author{D.\ E.\ Chang}
\author{F.\ H.\ L.\ Koppens}
\affiliation{ICFO-Institut de Ciencies Fotoniques, Mediterranean Technology Park, 08860 Castelldefels (Barcelona), Spain}
\author{F.\ J.\ Garc{\' i}a de Abajo}
\affiliation{ICFO-Institut de Ciencies Fotoniques, Mediterranean Technology Park, 08860 Castelldefels (Barcelona), Spain}
\affiliation{ICREA - Instituci— Catalana de Recerca i Estudis Avanats, Barcelona, Spain}
\author{M.\ D.\ Lukin}
\affiliation{Department of Physics, Harvard University, Cambridge, MA 02138, USA}

\pacs{78.67.Wj, 73.20.Mf, 42.50.-p, 42.65.Hw}
\date{\today}

\begin{abstract}
We show  that it is possible to realize significant nonlinear optical interactions at the few photon level in graphene nanostructures. Our approach takes advantage of the electric field enhancement associated with the strong confinement of graphene plasmons and the large intrinsic nonlinearity of graphene.    Such a system could  provide a powerful platform for quantum nonlinear optical control of light.  As an example, we consider an integrated optical device that exploits this large nonlinearity to realize a single photon switch.
\end{abstract}
\maketitle

\emph{Introduction} -
Nonlinear optical processes find ubiquitous use in modern
scientific and technological applications, facilitating diverse phenomena  like optical modulation and switching, spectroscopy, and frequency
conversion~\cite{boyd_nonlinear_2003}. A long-standing goal has been to realize nonlinear effects at progressively lower powers, which is difficult given the small nonlinear coefficients of bulk optical materials. The ultimate limit is  that of single-photon nonlinear optics, where individual photons strongly interact with each other. Realization of such nonlinear processes would not only facilitate peak performance of classical nonlinear devices, but also create a unique resource for implementation of  quantum networks~\cite{kimble_quantum_2008} and other applications that rely on the generation and manipulation of non-classical light.

One approach to reach the quantum regime involves coupling the light to individual quantum emitters to take advantage of their intrinsically nonlinear electronic spectrum \cite {duan_robust_2008,kimble_quantum_2008}. While a number of remarkable phenomena have been demonstrated with these systems \cite{Haroche13}, their realization remains a challenging task.   Specifically, in contrast to conventional bulk nonlinear systems, coherent single quantum emitters are generally unable to operate under ambient conditions, suffer from relatively slow operating speeds, are prone to strong decoherence in solid-state environments, and have limited tunability of their properties.

Fueled by these considerations, there has been renewed interest in nonlinear optical materials that can reach the quantum regime~\cite{Matsuda_fibreQNL_2009,mabuchi_qubit_2011,ferretti_single-photon_2012}. In particular, recent experiments demonstrated the realization of a quantum nonlinear medium, featuring  single photon blockade \cite{Peyronel12} and conditional two-photon phase shifts \cite{Peyronel13}, in a cold, dense gas of strongly interacting atoms. The essence of these approaches is that the interaction probability for two photons becomes substantial if the photons are confined to a sufficiently small mode volume of the nonlinear medium for sufficiently long times.  Motivated by these recent developments, in this Letter we explore the potential for using nanoscale surface plasmon excitations in graphene for quantum nonlinear optics.  Graphene, a single atomic layer of carbon atoms, has attracted tremendous interest for its unique electronic, mechanical, and quantum transport properties~\cite{geim_rise_2007,castro_neto_electronic_2009,nair_fine_2008}.  
Recently it has also been realized that the unique properties of graphene have a strong effect on the guided electromagnetic surface waves in the form of surface plasmons~(SPs) \cite{wunsch_spDisp_2006,mikhailov_new_2007,jablan_plasmonics_2009}.
 In particular, recent theoretical~\cite{mikhailov_new_2007,jablan_plasmonics_2009,koppens_graphene_2011,Nikitin12} and experimental~\cite{fei_gate-tuning_2012,chen_optical_2012,Fang13} results indicate that graphene plasmons can be confined to volumes millions of times smaller than in free space.  We show that under realistic conditions, this field confinement enables deterministic interaction between two plasmons~(\textit{i.e.}, photons) over picosecond time scales as illustrated in Fig. 1ab, which is much shorter than the anticipated  plasmon lifetime \cite{Principi13}. We show how one can take advantage of this interaction to realize a single photon switch (Fig. 1cd) and produce non-classical light. 

\begin{figure}[htbp]
\begin{center}
\includegraphics[width=0.49 \textwidth]{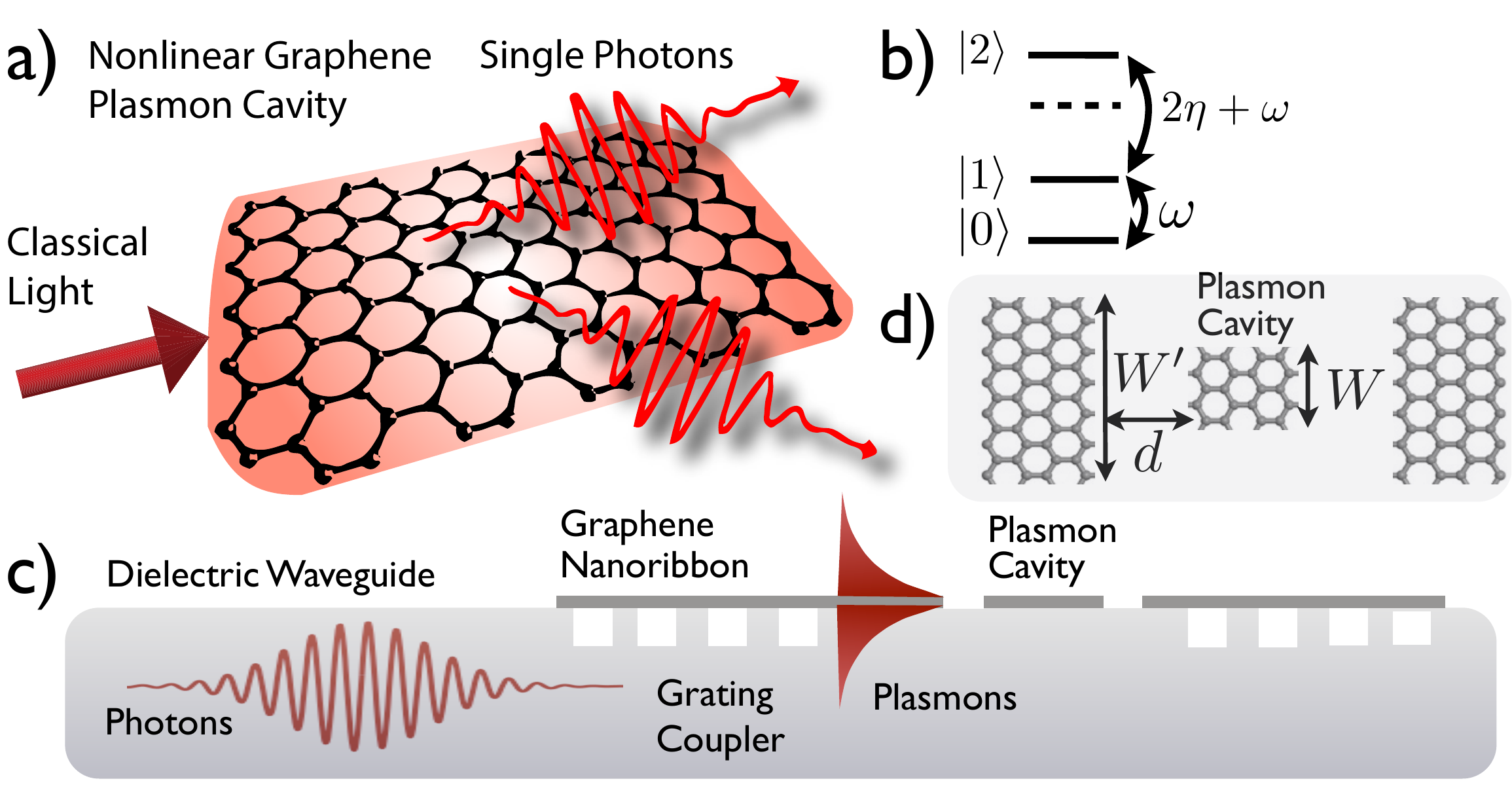}
\caption{a) A doped graphene disk confines photons as plasmons to mode volumes millions of times smaller than free space.  b) This induces a large dispersive nonlinearity $\eta$ (defined in Eq.\ \ref{eqn:etaq}) so that only a single photon can resonantly excite the cavity.  c) Integrated nonlinear optical circuit for using the graphene plasmon cavity to realize a single photon switch.  First the photons are converted into planar plasmons of a graphene waveguide via a grating, then they couple to the plasmon cavity, after which they are converted back into waveguide photons.  For the frequencies we consider the waveguide and grating could be fabricated from etched Si.  d) Top down view of the plasmon cavity from (c) showing the width $W$ of the plasmon cavity, the width $W'$ the graphene nanoribbon, and the spacing $d$ between the cavity and nanoribbon.
}
\label{fig:GQNL1}
\end{center}
\end{figure}

Through electrostatic gating, it is possible to introduce a net carrier concentration, which shifts the Fermi energy $\hbar\omega_F$ away from the Dirac point to a non-zero value. The in-plane conductivity of graphene is well-approximated by the
expression
$\sigma(\omega)\approx\frac{ie^2}{\pi\hbar}\frac{\omega_F}{\omega+i
\gamma}$ at frequencies below twice the Fermi frequency
$\omega<2\omega_F$~\cite{falkovsky_optical_2008}, which describes a
Drude-like response of electrons within a single band. In
realistic systems the conductivity will also have a small term
$\gamma$ describing dissipation due impurity or
phonon-mediated scattering. There are two limits on the existence of low-loss SP modes in graphene.  First, at frequencies $\omega>2\omega_F$,
graphene suffers from strong inter-band absorption
\cite{jablan_plasmonics_2009,koppens_graphene_2011}.  Second for frequencies above the  optical phonon frequency $\hbar \omega_{op} \approx 0.2$ eV, there is additional loss due to scattering into optical phonons  \cite{jablan_plasmonics_2009,Yan2013};
although narrow plasmons above $ \omega_{op}$ have  recently been observed in graphene nanorings \cite{Fang13}.
To minimize the losses we focus on the regime where the frequencies fall below $2\omega_F$ and $\omega_{op}$.  In this regime, we can approximate $\gamma = e v_F^2/\mu\, \hbar \omega_F$ where $\mu$ is the mobility \cite{Novoselov05}.  The ability to tune $\omega_F$, and consequently the optical properties,
through electrostatic gating makes graphene unique compared to normal metals.

Like in noble-metal plasmonics~\cite{Barnes03}, the free nature of charge carriers described by the Drude
response gives rise to SP modes in
graphene~\cite{mikhailov_new_2007,jablan_plasmonics_2009}.
At first order in $k_{sp}/k_F$ the SP dispersion is given by
\be \label{eqn:sp_disp}
\omega_{sp}^2= \frac{e^2\omega_F}{2 \pi \epsilon_0  \hbar}   k_{sp} \approx 4.2\, \omega_F\, v_F\, k_{sp}
\ee
where $v_F \approx 10^{6}$ m/s is the Fermi velocity \cite{wunsch_spDisp_2006}.  This dispersion relation implies a remarkable reduction of the SP wavelength compared to the free space wavelength $\lambda_0=2\pi c/\omega_{sp}$, as $\lambda_{sp}/ \lambda_0 \sim v_F/c\sim 3 \cdot 10^{-3}$.  Thus, the smallest possible mode volume of a graphene SP resonator, $V\sim \lambda_{sp}^3$, can be~$\sim 10^7$ times smaller than in free space \cite{koppens_graphene_2011}.

\emph{Nonlinear plasmonics in graphene}~\label{sec:graphene} -
To describe the nonlinear properties of the plasmons we employ the semiclassical Maxwell-Boltzmann  equations (MBE).  This is a good approximation when the plasmon momentum is much less than the Fermi momentum and the plasmon properties are dominated by intra-band transitions.
The distribution function $f(\bm{x},\bm{k},t)$ for an electron at in-plane position $\bm{x}$ and with Bloch momentum $\bm{k}$ evolves under the Maxwell-Boltzmann equation as
\be
\del_t f +v_F\hat{k} \cdot \del_{\bm{x}} f+ e \del_{\bm{x}} \varphi \cdot \del_{\bm{k}} f = 0,
\ee
where the electostatic potential $\phi(\bm{x},z,t)$ satisfies Poisson's equation $\nabla^2 \phi = en\delta(z)/\epsilon_0 \epsilon$. Here $z$ is the out-of-plane coordinate and $n=\int\,d\bm{k}\,f$ is the 2D electron density. For weak excitations of the electron distribution, the term $\partial_{\bm{k}}f$ in the Maxwell-Boltzmann equation can be replaced by the equilibrium value $\partial_{\bm{k}}f^{(0)}$, yielding a linear equation supporting SPs with the dispersion given in Eq.~(\ref{eqn:sp_disp}) and an electrostatic wave given by $E =-\nabla\phi \propto \delta n \sin(kx -\omega t)$.

For sufficiently large density perturbations $\delta n$, the nonlinear interaction between the non-equilibrium distribution $\partial_{\bm{k}}f$ and potential must be accounted for. This effect can be interpreted as a backaction induced by the electrostatic wave on the electrons via a ponderomotive force $F_p \sim \del_{x} E^2 \sim k \delta n^2 \sin 2 k x$, which grows with the amplitude of the SPs.  This nonlinear force directly excites a second plasmon wave at wavevector $2k$ and frequency $2 \omega$, i.e.~second harmonic generation, and gives rise to the second order conductivity calculated in Ref.~\cite{mikhailov2011}. We show~(see supplementary material \cite{supp}) that this leads to a nonlinear shift at the original wavevector $k$ and frequency $\omega$, with an effective third order conductivity for the SPs given by 
\be
\sigma^{(3)}(k_{sp},\omega)
= -i \frac{3 \pi  }{4}  \frac{ v_F^4}{\omega_F^3 } \frac{\epsilon_0^2}{\hbar\, \omega}.
\ee
This result differs from the nonlinear conductivity as seen by free-space light normally incident on a graphene sheet, where one finds that $\sigma^{(3)} \sim 1/\omega^3$ \cite{mikhailov2008}. Remarkably, as we discuss next, the tight confinement of SPs in graphene implies that the fields associated with even single quantized SPs are strong enough that nonlinear effects are observable.

\emph{Nonlinear Graphene~Plasmon Cavity}~\label{sec:graphene}-
Anticipating the large strength of nonlinear interactions at the level of single SPs in nanoscale graphene resonators, we are motivated to introduce a quantum description of such a system. We write the Hamiltonian  as $H=H_0+H_c$, where $H_0$ characterizes the excitation spectrum of the graphene resonator, and $H_c$ describes an external coupling to the resonator (as in Fig.\ 1cd), which allows one to probe the resonator properties or utilize the nonlinearities for applications such as a single-photon transistor.

\begin{figure}[htbp]
\begin{center}
\includegraphics[width=0.49 \textwidth]{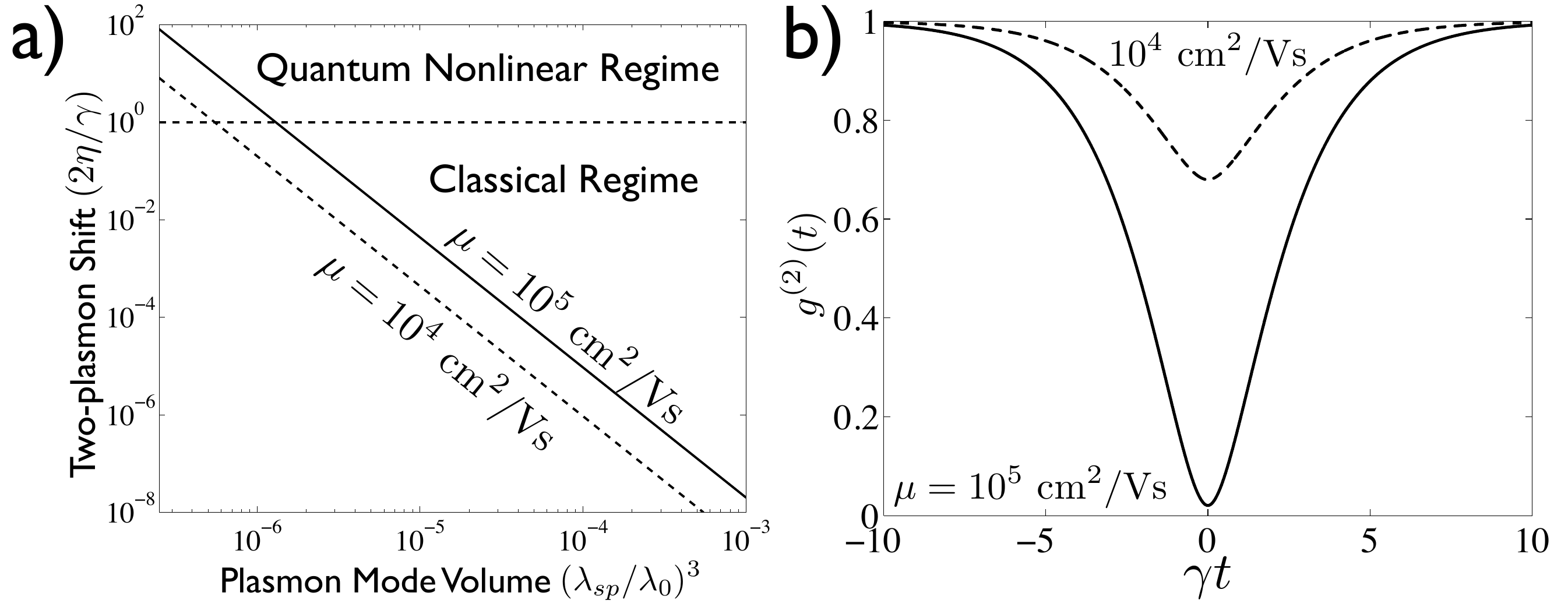}
\caption{a)  Nonlinear shift (calculated from Eq.~\ref{eqn:etaq}) for the fundamental mode relative to the plasmon linewidth with decreasing mode volume $V_0=(\lambda_{sp}/\lambda_0)^3$.  Here we take the linewidth as $\gamma = e v_F^2/\mu\, \hbar \omega_F$ with the Fermi energy $\hbar \omega_F = 0.2$~eV and a mobility of $\mu=10^5(10^4)$~cm$^2$/Vs corresponding to a quality factor of roughly 600(60).   
b)  $g^{(2)}(t)$ for the graphene plasmon cavity driven by a weak coherent state for $\hbar \omega_{sp} = 0.2$ eV and two different mobility.  $g^{(2)}(0)<1$ indicates a transition to an effective two-level system as illustrated schematically in Fig.\ 1ab. }
\label{fig:GQNL2}
\end{center}
\end{figure}

We first consider the intrinsic properties of the resonator given by $H_0$. Considering the fundamental SP mode of the resonator with corresponding annihilation operator $a_q$ and number operator $n_q=a^{\dagger}_qa_q$, the effective Hamiltonian $H_0$ is given by~\cite{Denardo88,Gervasoni03,supp}
\be \label{eqn:Ham}
H_0=\left(\omega_q-i\kappa/2+\eta_q(n_q-1)\right)n_q.
\ee
This Hamiltonian describes the quantum analog of a cavity exhibiting an intensity-dependent refractive index, where the  resonance frequency $\omega_q+\eta_q(n_q-1)$ shifts depending on the intra-cavity photon number. Here we have also included the total cavity linewidth $\kappa=\kappa_{ex}+\gamma$ into the cavity description which includes  the intrinsic losses $\gamma$
and radiative losses of the cavity into other optical or plasmonic modes, given by $\kappa_{ex}$. For graphene, the nonlinear interaction strength is given by \cite{supp}
\be \label{eqn:etaq}
\eta_q =\frac{7\, \pi\, \omega_q}{64\, A\ k_F^2} \sqrt{\frac{  q^3}{2 \alpha_g k_F^3} },
\ee
where $\alpha_g \equiv e^2/4 \pi \epsilon_0\, \hbar\, v_F \approx 2$ and $A$ is the mode area of the resonator, which can be given by $A=\lambda_{sp}^2/4$ for a diffraction-limited structure.  The $\eta_q\propto A^{-1}$ scaling reflects that the field intensity of a single SP grows inversely with its confinement.  For such small structures one might expect that quantum size effects become important; however, as shown in Ref.\ \cite{Javier12}, for graphene nanostructures larger than $\sim10-20$ nm the use of the bulk dielectric response is a valid approximation.

At the quantum level, the interaction parameter $2 \eta_q$ indicates the additional energy cost to excite two versus one photon in the cavity, as can be seen in the cavity excitation spectrum~(Fig.\ \ref{fig:GQNL1}b). When $2 \eta \gg \kappa$, the graphene sheet behaves as a two-level atom because it can only resonantly absorb a single photon as illustrated in Fig.\ \ref{fig:GQNL1}a; thus we describe this as the quantum nonlinear regime.  The ratio $2 \eta_q/\kappa$ is then a good measure of the quality of the cavity as a quantum emitter.  Fig.~\ref{fig:GQNL2}b shows $2 \eta_q/\gamma$ for the fundamental mode with decreasing mode volume (assuming mobilities of $10^5$ and $10^4$ cm$^2$/Vs), where we see that this ratio can be as large as 100. The parameter $\eta/\kappa\propto Q/A$, where $Q$ is the quality factor of the resonator. 

The enabling mechanism for a two-level atom to be useful for quantum information processing is that it can only emit single photons at a time. This can be characterized by the second order correlation function of the emitted light, which is identical to that of the cavity mode, $g^{(2)}(t) = \langle a^\dagger(\tau) a^\dagger(t+\tau) a(t+\tau)) a(\tau) \rangle/ \langle a^\dagger(\tau) a(\tau) \rangle$. For a stationary process, $g^{(2)}(0)<1$ indicates non-classical ``anti-bunching'' and approaches $g^{(2)}(0)=0$ in the limit of an ideal two-level emitter.  We consider the case where the resonator is driven by an external laser from the side and emission is collected from a different direction.  In the limit of weak driving we find that
\be \label{eqn:g2}
g^{(2)}(0) = \frac{\kappa^2}{4 \eta^2 + \kappa^2},
\ee
thus establishing $\eta \lesssim \kappa$ as the regime where quantum properties become observable. In Fig.~\ref{fig:GQNL2}b we take $\kappa_{ex}=0$ and we see that, for the largest nonlinearities, $g^{(2)}<1$ can be readily observed for high mobility graphene.

\emph{Efficient coupling and a single-photon switch}~\label{sec:coupling}- 
In order to exploit the large nonlinearity of graphene, we need an efficient method to convert SPs into external optical modes on time scales short compared to the intrinsic losses. Specifically, one needs that the total linewidth $\kappa=\kappa_{ex}+\gamma$ contains a large component $\kappa_{ex}$ that goes into desirable external channels compared to the intrinsic losses $\gamma$. One approach is to use the direct dipolar emission of the cavity into free space radiation.  For the square cavities described above, the dipole moment is given by $p = 2 \,e\, k_F^2/k_{sp}^3 $ which gives a decay rate into radiation of
\be \label{eqn:kappaEx}
\kappa_{ex} =\frac{k_0^3\, p^2}{3 \pi \epsilon_0 \hbar}= \frac{16\, \alpha_g}{3}\frac{ k_F^3}{k_{sp}^3} V_0\, \omega_F
\ee
where $V_0 \equiv (\lambda_{sp}/\lambda_0)^3$.
For cavities in the quantum nonlinear regime, this is a small contribution to the total losses; thus, 
a more practical approach is needed for coupling the photons to free space.

\begin{figure*}[thbp]
\begin{center}
\includegraphics[width=0.9 \textwidth]{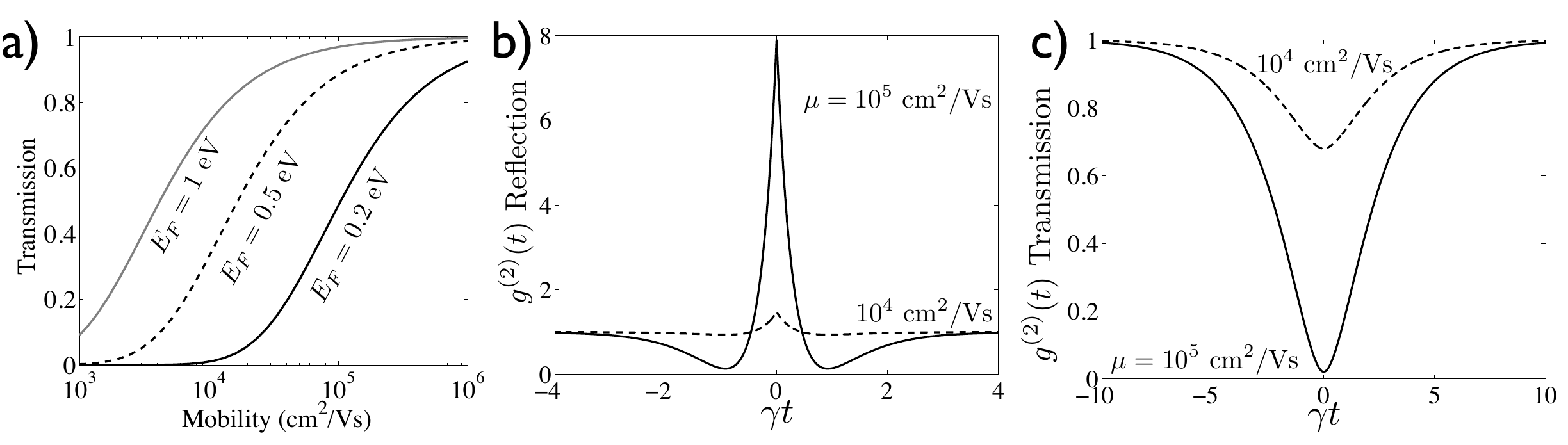}
\caption{a) Single photon transmission through the device depicted in Fig.\ 1cd. We take $\xi= k_0/2$ and $P\gg1$ so the only losses are in the nanoribbons.  The plasmon frequency is 0.2 eV and we assume the decay rate $\gamma$ is dominated by impurity scattering.   The three curves are for a fixed plasmon frequency with increasing Fermi energy, which increases the spatial propagation length of the plasmons.  b) Bunching in reflection for two incident photons from the left with $\hbar \omega_{sp}=0.2$ eV, $E_F = 0.23$ eV, $P=2$, and mobility $\mu=10^4(10^5)$~cm$^2$/V$\cdot$s (dashed(solid)) corresponding to a lifetime of $0.2(2)$ ps  and a cavity  quality factor of 60(600). e) Antibunching in transmission for $P=0.1$ with other parameters as in (a).  }
\label{fig:GQNL3}
\end{center}
\end{figure*}

We envision a two-step process illustrated in Fig.~\ref{fig:GQNL1}cd: first a waveguide photon is converted into  the planar plasmon of a graphene waveguide via a  dielectric grating, then this plasmon can tunnel directly into the nonlinear cavity.
We first consider the direct coupling between the cavity and the bulk plasmons.  We take the cavity of width $W$ to be separated a distance $d$ from a long nanoribbon of width $W'$ as shown in Fig.\ 1d.  For $d\gg W,\lambda_{sp}$ the coupling is dipolar and small, which allows us to calculate the decay of the fundamental cavity mode into the nanoribbon via Fermi's golden rule \cite{supp}
\be \label{eqn:kappaq}
\kappa_{c-r}= \frac{32}{\pi^2} \frac{k_F^r}{k_F^c} \frac{ W\, \omega }{k_{sp}^* k_{sp}^4 d^6}
\ee
where $k_F^{r,c}$ is the Fermi wavevector in the nanoribbon ($r$) and cavity ($c$) and $k_{sp}^*$ is the wavevector for the nanoribbon plasmon that is resonant with the cavity mode.
The cavity can be efficiently controlled through the nanoribbon by operating at a distance $d$ such that this decay is the dominant loss channel for the cavity.

Once the plasmon is in the nanoribbon it still remains to out-couple it to the waveguide.  Due to the large mismatch in wavevectors, $k_{sp}/k_0 \sim c/v_F$, the bare coupling of the plasmons to the waveguide mode will be very small.  Achieving efficient conversion between conventional optics and plasmons (in graphene and noble metals) is an area of active research with several different approaches being pursued \cite{Bao12,Ruppert10,Zhan12,Bludov12,Aires12,Nikitin12b,Zhu13}.  
We consider the conceptually simplest solution, which is to directly fabricate a dielectric grating to enable momentum conservation, as recently demonstrated in Ref.~\cite{Zhu13}.  
We take a single-mode dielectric slab waveguide in vacuum coupled to a graphene nanoribbon via the dielectric grating.  For the frequencies we consider here one could fabricate the waveguide and grating by etching Si.   For parallel propagation, the grating wavector $k_g$ should be given by $k_g= k_{sp}-k_0$.  This  geometry can be analyzed via coupled mode theory and optimized as a function of the slab thickness \cite{Snyder83}.  Taking the grating profile to be of the form $\epsilon_g(x)= \delta \epsilon \cos k_g x$ gives the power conversion for weak losses between the waveguide and plasmon mode as $\cos^2 (\xi x)$ where $\xi$ is
spatial coupling between the TM mode of the waveguide and nanoribbon 
\be
\xi \approx  \sqrt{\frac{W}{W'}}\, \delta \epsilon\, e^{-\gamma_\perp h}  k_0
\ee
Here $W'>W$,  $\gamma_\perp^2 = \beta^2 - k_0^2$ is the transverse wavevector of the slab mode, $\beta$ is the longitudinal wavevector, and $h$ is the distance between the slab and the graphene.   Because the factor in $\xi$ in front of $k_0$ is order unity, the plasmon conversion for a weak grating is limited to distances $\sim \lambda_0 \gg \lambda_{sp}$.  As a result the spatial decay rate of the plasmons must be much larger than $k_0$ to achieve efficient conversion.  When losses are dominated by impurity scattering the spatial decay rate is given by $\gamma\, k_{sp}/\omega_{sp} \approx e v_F \,\hbar\, \omega_{sp}/2\, \mu\, E_F^2$, which decreases with Fermi energy.  Fig.~\ref{fig:GQNL3}a  shows the transmission of a single photon through the geometry displayed in Fig.~\ref{fig:GQNL1}cd.

The device depicted in Fig.~\ref{fig:GQNL3}ab can be used as a nonlinear single-photon switch. To characterize this process, it is first necessary to understand how an input field through the waveguide is transformed upon interacting with the nonlinear resonator, which can be done through an input-output formalism. In the case of Fig.~\ref{fig:GQNL3}ab of a resonator equally coupled to two waveguides, the resonator evolves under the incoming fields of the left- and right-going modes under the Hamiltonian $H_c=\sqrt{\kappa_{ex}}(a^{r}_{in}+a^{l}_{in})a^{\dagger}+h.c.$, while the output fields are given by $a^{r(l)}_{out}=a^{r(l)}_{in}+i\sqrt{\kappa_{ex}}a$.

This one dimensional model has been solved exactly for the case of one and two resonant photons input from a single direction in the waveguide \cite{Liao10}.  The response is characterized by the effective Purcell factor $P= \kappa_{ex}/\gamma$, which measures the fraction of cavity emission into the waveguide, and the normalized nonlinearity $\tilde{\eta}= \eta/\kappa$.  The transmission $t$ and reflection $r$ coefficients for a single photon incident on resonance with the cavity are given by $t=-P/(1+P)$ and $r=1/(1+P)$.  The two photon response, however, is modified by the nonlinearity.  For example two photons at frequency $\omega_{sp}$ will be blocked from entering the cavity due to the nonlinearity.  This leads to antibunching in the transmission and bunching in the reflection as shown in Figures \ref{fig:GQNL3}bc.  The suppression in the transmission scales as $\tilde{\eta}^2$ similarly to Eq.~\ref{eqn:g2}, while the bunching in reflection scales as $P^4$ for $\tilde{\eta} \gg P \gg 1$ \cite{Liao10}.  Fig.~\ref{fig:GQNL3}c shows that such a device realizes a single photon transistor where one control photon can block several signal photons from propagating through the cavity for a time given by the inverse cavity lifetime. 


Experimental tests of these ideas require single photon detectors in the mid-infrared wavelength regime between 2 and 10 $\mu$m.  While this is challenging to realize, recent work on superconducting nanowire single photon detectors (SNSPD) and transition edge sensors (TES) have achieved single photon detection in this regime \cite{Eisaman11,Marsili12,Karasik12}.  Alternatively, frequency up-conversion may allow efficient detection \cite{Eisaman11}.

Our analysis shows that graphene plasmonics may provide a powerful platform for the nonlinear quantum optical control of light.  Combined with the scalable fabrication of graphene this could allow the creation of complex quantum networks for many applications in quantum information and quantum simulation, as well as in classical nonlinear optics.  Such a system is ultimately limited either by the losses in graphene or the strength of the nonlinearity.  We estimate currently achievable quality factors for the plasmon cavity range from $10-10^3$; however, estimates of the ultimate limit to the graphene plasmon lifetime suggest that quality factors greater than $10^4$ are possible \cite{Principi13}.  To enhance the nonlinearity further hybrid structures can be envisioned if one could fabricate the structure on top of a strong nonlinear substrate.  

\emph{Acknowledgments--} We thank P.~Zoller for useful discussions. DEC acknowledges support from Fundacio Privada Cellex Barcelona.  We acknowledge support from the Harvard Center for Quantum Optics, NSF, ARO, CUA, AFOSR MURI and the Packard Foundation.

\bibliographystyle{apsrev-nourl}
\bibliography{graphene_NLO_bib}

\begin{thebibliography}{44}
\expandafter\ifx\csname natexlab\endcsname\relax\def\natexlab#1{#1}\fi
\expandafter\ifx\csname bibnamefont\endcsname\relax
  \def\bibnamefont#1{#1}\fi
\expandafter\ifx\csname bibfnamefont\endcsname\relax
  \def\bibfnamefont#1{#1}\fi
\expandafter\ifx\csname citenamefont\endcsname\relax
  \def\citenamefont#1{#1}\fi
\expandafter\ifx\csname url\endcsname\relax
  \def\url#1{\texttt{#1}}\fi
\expandafter\ifx\csname urlprefix\endcsname\relax\def\urlprefix{URL }\fi
\providecommand{\bibinfo}[2]{#2}
\providecommand{\eprint}[2][]{\url{#2}}

\bibitem[{\citenamefont{Boyd}(2003)}]{boyd_nonlinear_2003}
\bibinfo{author}{\bibfnamefont{R.~W.} \bibnamefont{Boyd}},
  \emph{\bibinfo{title}{Nonlinear optics}} (\bibinfo{publisher}{Academic
  Press}, \bibinfo{address}{San Diego}, \bibinfo{year}{2003}).

\bibitem[{\citenamefont{Kimble}(2008)}]{kimble_quantum_2008}
\bibinfo{author}{\bibfnamefont{H.~J.} \bibnamefont{Kimble}},
  \bibinfo{journal}{Nature} \textbf{\bibinfo{volume}{453}},
  \bibinfo{pages}{1023} (\bibinfo{year}{2008}).

\bibitem[{\citenamefont{Duan and Monroe}(2008)}]{duan_robust_2008}
\bibinfo{author}{\bibfnamefont{L.-M.} \bibnamefont{Duan}} \bibnamefont{and}
  \bibinfo{author}{\bibfnamefont{C.}~\bibnamefont{Monroe}}, in
  \emph{\bibinfo{booktitle}{Advances In Atomic, Molecular, and Optical
  Physics}}, edited by
  \bibinfo{editor}{\bibfnamefont{E.}~\bibnamefont{Arimondo}},
  \bibinfo{editor}{\bibfnamefont{P.~R.} \bibnamefont{Berman}},
  \bibnamefont{and} \bibinfo{editor}{\bibfnamefont{C.~C.} \bibnamefont{Lin}}
  (\bibinfo{publisher}{Academic Press}, \bibinfo{year}{2008}),
  vol.~\bibinfo{volume}{55}, pp. \bibinfo{pages}{419--463}.

\bibitem[{\citenamefont{Haroche}(2013)}]{Haroche13}
\bibinfo{author}{\bibfnamefont{S.}~\bibnamefont{Haroche}},
  \bibinfo{journal}{Rev. Mod. Phys.} \textbf{\bibinfo{volume}{85}},
  \bibinfo{pages}{1083} (\bibinfo{year}{2013}).

\bibitem[{\citenamefont{Matsuda et~al.}(2009)\citenamefont{Matsuda, Shimizu,
  Mitsumori, Kosaka, and Edamatsu}}]{Matsuda_fibreQNL_2009}
\bibinfo{author}{\bibfnamefont{N.}~\bibnamefont{Matsuda}},
  \bibinfo{author}{\bibfnamefont{R.}~\bibnamefont{Shimizu}},
  \bibinfo{author}{\bibfnamefont{Y.}~\bibnamefont{Mitsumori}},
  \bibinfo{author}{\bibfnamefont{H.}~\bibnamefont{Kosaka}}, \bibnamefont{and}
  \bibinfo{author}{\bibfnamefont{K.}~\bibnamefont{Edamatsu}},
  \bibinfo{journal}{Nat. Phot.} \textbf{\bibinfo{volume}{3}},
  \bibinfo{pages}{95} (\bibinfo{year}{2009}).

\bibitem[{\citenamefont{Mabuchi}(2012)}]{mabuchi_qubit_2011}
\bibinfo{author}{\bibfnamefont{H.}~\bibnamefont{Mabuchi}},
  \bibinfo{journal}{Phys. Rev. A} \textbf{\bibinfo{volume}{85}},
  \bibinfo{pages}{015806} (\bibinfo{year}{2012}).

\bibitem[{\citenamefont{Ferretti and
  Gerace}(2012)}]{ferretti_single-photon_2012}
\bibinfo{author}{\bibfnamefont{S.}~\bibnamefont{Ferretti}} \bibnamefont{and}
  \bibinfo{author}{\bibfnamefont{D.}~\bibnamefont{Gerace}},
  \bibinfo{journal}{Phys. Rev. B} \textbf{\bibinfo{volume}{85}},
  \bibinfo{pages}{033303} (\bibinfo{year}{2012}).

\bibitem[{\citenamefont{Peyronel et~al.}(2012)\citenamefont{Peyronel,
  Firstenberg, Liang, Hofferberth, Gorshkov, Pohl, Lukin, and
  Vuleti{\'c}}}]{Peyronel12}
\bibinfo{author}{\bibfnamefont{T.}~\bibnamefont{Peyronel}},
  \bibinfo{author}{\bibfnamefont{O.}~\bibnamefont{Firstenberg}},
  \bibinfo{author}{\bibfnamefont{Q.-Y.} \bibnamefont{Liang}},
  \bibinfo{author}{\bibfnamefont{S.}~\bibnamefont{Hofferberth}},
  \bibinfo{author}{\bibfnamefont{A.~V.} \bibnamefont{Gorshkov}},
  \bibinfo{author}{\bibfnamefont{T.}~\bibnamefont{Pohl}},
  \bibinfo{author}{\bibfnamefont{M.~D.} \bibnamefont{Lukin}}, \bibnamefont{and}
  \bibinfo{author}{\bibfnamefont{V.}~\bibnamefont{Vuleti{\'c}}},
  \bibinfo{journal}{Nature} \textbf{\bibinfo{volume}{488}}, \bibinfo{pages}{57}
  (\bibinfo{year}{2012}).

\bibitem[{\citenamefont{Peyronel et~al.}(2013)\citenamefont{Peyronel,
  Firstenberg, Liang, Hofferberth, Gorshkov, Lukin, and
  Vuleti{\'c}}}]{Peyronel13}
\bibinfo{author}{\bibfnamefont{T.}~\bibnamefont{Peyronel}},
  \bibinfo{author}{\bibfnamefont{O.}~\bibnamefont{Firstenberg}},
  \bibinfo{author}{\bibfnamefont{Q.~Y.} \bibnamefont{Liang}},
  \bibinfo{author}{\bibfnamefont{S.}~\bibnamefont{Hofferberth}},
  \bibinfo{author}{\bibfnamefont{A.~V.} \bibnamefont{Gorshkov}},
  \bibinfo{author}{\bibfnamefont{M.~D.} \bibnamefont{Lukin}}, \bibnamefont{and}
  \bibinfo{author}{\bibfnamefont{V.}~\bibnamefont{Vuleti{\'c}}},
  \bibinfo{journal}{Nature, in press.}  (\bibinfo{year}{2013}).

\bibitem[{\citenamefont{Geim and Novoselov}(2007)}]{geim_rise_2007}
\bibinfo{author}{\bibfnamefont{A.~K.} \bibnamefont{Geim}} \bibnamefont{and}
  \bibinfo{author}{\bibfnamefont{K.~S.} \bibnamefont{Novoselov}},
  \bibinfo{journal}{Nat. Mater.} \textbf{\bibinfo{volume}{6}},
  \bibinfo{pages}{183} (\bibinfo{year}{2007}).

\bibitem[{\citenamefont{Castro~Neto et~al.}(2009)\citenamefont{Castro~Neto,
  Guinea, Peres, Novoselov, and Geim}}]{castro_neto_electronic_2009}
\bibinfo{author}{\bibfnamefont{A.~H.} \bibnamefont{Castro~Neto}},
  \bibinfo{author}{\bibfnamefont{F.}~\bibnamefont{Guinea}},
  \bibinfo{author}{\bibfnamefont{N.~M.~R.} \bibnamefont{Peres}},
  \bibinfo{author}{\bibfnamefont{K.~S.} \bibnamefont{Novoselov}},
  \bibnamefont{and} \bibinfo{author}{\bibfnamefont{A.~K.} \bibnamefont{Geim}},
  \bibinfo{journal}{Rev. Mod. Phys.} \textbf{\bibinfo{volume}{81}},
  \bibinfo{pages}{109} (\bibinfo{year}{2009}).

\bibitem[{\citenamefont{Nair et~al.}(2008)\citenamefont{Nair, Blake,
  Grigorenko, Novoselov, Booth, Stauber, Peres, and Geim}}]{nair_fine_2008}
\bibinfo{author}{\bibfnamefont{R.~R.} \bibnamefont{Nair}},
  \bibinfo{author}{\bibfnamefont{P.}~\bibnamefont{Blake}},
  \bibinfo{author}{\bibfnamefont{A.~N.} \bibnamefont{Grigorenko}},
  \bibinfo{author}{\bibfnamefont{K.~S.} \bibnamefont{Novoselov}},
  \bibinfo{author}{\bibfnamefont{T.~J.} \bibnamefont{Booth}},
  \bibinfo{author}{\bibfnamefont{T.}~\bibnamefont{Stauber}},
  \bibinfo{author}{\bibfnamefont{N.~M.~R.} \bibnamefont{Peres}},
  \bibnamefont{and} \bibinfo{author}{\bibfnamefont{A.~K.} \bibnamefont{Geim}},
  \bibinfo{journal}{Science} \textbf{\bibinfo{volume}{320}},
  \bibinfo{pages}{1308} (\bibinfo{year}{2008}).

\bibitem[{\citenamefont{Wunsch et~al.}(2006)\citenamefont{Wunsch, Stauber,
  Sols, and Guinea}}]{wunsch_spDisp_2006}
\bibinfo{author}{\bibfnamefont{B.}~\bibnamefont{Wunsch}},
  \bibinfo{author}{\bibfnamefont{T.}~\bibnamefont{Stauber}},
  \bibinfo{author}{\bibfnamefont{F.}~\bibnamefont{Sols}}, \bibnamefont{and}
  \bibinfo{author}{\bibfnamefont{F.}~\bibnamefont{Guinea}},
  \bibinfo{journal}{New J. Phys.} \textbf{\bibinfo{volume}{8}},
  \bibinfo{pages}{318} (\bibinfo{year}{2006}).

\bibitem[{\citenamefont{Mikhailov and Ziegler}(2007)}]{mikhailov_new_2007}
\bibinfo{author}{\bibfnamefont{S.~A.} \bibnamefont{Mikhailov}}
  \bibnamefont{and} \bibinfo{author}{\bibfnamefont{K.}~\bibnamefont{Ziegler}},
  \bibinfo{journal}{Phys. Rev. Lett.} \textbf{\bibinfo{volume}{99}},
  \bibinfo{pages}{016803} (\bibinfo{year}{2007}).

\bibitem[{\citenamefont{Jablan et~al.}(2009)\citenamefont{Jablan, Buljan, and
  Solja{\v c}i{\' c}}}]{jablan_plasmonics_2009}
\bibinfo{author}{\bibfnamefont{M.}~\bibnamefont{Jablan}},
  \bibinfo{author}{\bibfnamefont{H.}~\bibnamefont{Buljan}}, \bibnamefont{and}
  \bibinfo{author}{\bibfnamefont{M.}~\bibnamefont{Solja{\v c}i{\' c}}},
  \bibinfo{journal}{Phys. Rev. B} \textbf{\bibinfo{volume}{80}},
  \bibinfo{pages}{245435} (\bibinfo{year}{2009}).

\bibitem[{\citenamefont{Koppens et~al.}(2011)\citenamefont{Koppens, Chang, and
  García~de Abajo}}]{koppens_graphene_2011}
\bibinfo{author}{\bibfnamefont{F.~H.~L.} \bibnamefont{Koppens}},
  \bibinfo{author}{\bibfnamefont{D.~E.} \bibnamefont{Chang}}, \bibnamefont{and}
  \bibinfo{author}{\bibfnamefont{F.~J.} \bibnamefont{García~de Abajo}},
  \bibinfo{journal}{Nano Lett.} \textbf{\bibinfo{volume}{11}},
  \bibinfo{pages}{3370} (\bibinfo{year}{2011}).

\bibitem[{\citenamefont{Nikitin
  et~al.}(2012{\natexlab{a}})\citenamefont{Nikitin, Guinea, Garcia-Vidal, and
  Martin-Moreno}}]{Nikitin12}
\bibinfo{author}{\bibfnamefont{A.~Y.} \bibnamefont{Nikitin}},
  \bibinfo{author}{\bibfnamefont{F.}~\bibnamefont{Guinea}},
  \bibinfo{author}{\bibfnamefont{F.~J.} \bibnamefont{Garcia-Vidal}},
  \bibnamefont{and}
  \bibinfo{author}{\bibfnamefont{L.}~\bibnamefont{Martin-Moreno}},
  \bibinfo{journal}{Phys. Rev. B} \textbf{\bibinfo{volume}{85}},
  \bibinfo{pages}{081405} (\bibinfo{year}{2012}{\natexlab{a}}).

\bibitem[{\citenamefont{Fei et~al.}(2012)\citenamefont{Fei, Rodin, Andreev,
  Bao, {McLeod}, Wagner, Zhang, Zhao, Thiemens, Dominguez
  et~al.}}]{fei_gate-tuning_2012}
\bibinfo{author}{\bibfnamefont{Z.}~\bibnamefont{Fei}},
  \bibinfo{author}{\bibfnamefont{A.~S.} \bibnamefont{Rodin}},
  \bibinfo{author}{\bibfnamefont{G.~O.} \bibnamefont{Andreev}},
  \bibinfo{author}{\bibfnamefont{W.}~\bibnamefont{Bao}},
  \bibinfo{author}{\bibfnamefont{A.~S.} \bibnamefont{{McLeod}}},
  \bibinfo{author}{\bibfnamefont{M.}~\bibnamefont{Wagner}},
  \bibinfo{author}{\bibfnamefont{L.~M.} \bibnamefont{Zhang}},
  \bibinfo{author}{\bibfnamefont{Z.}~\bibnamefont{Zhao}},
  \bibinfo{author}{\bibfnamefont{M.}~\bibnamefont{Thiemens}},
  \bibinfo{author}{\bibfnamefont{G.}~\bibnamefont{Dominguez}},
  \bibnamefont{et~al.}, \bibinfo{journal}{Nature}
  \textbf{\bibinfo{volume}{487}}, \bibinfo{pages}{82} (\bibinfo{year}{2012}).

\bibitem[{\citenamefont{Chen et~al.}(2012)\citenamefont{Chen, Badioli,
  Alonso-González, Thongrattanasiri, Huth, Osmond, Spasenović, Centeno,
  Pesquera, Godignon et~al.}}]{chen_optical_2012}
\bibinfo{author}{\bibfnamefont{J.}~\bibnamefont{Chen}},
  \bibinfo{author}{\bibfnamefont{M.}~\bibnamefont{Badioli}},
  \bibinfo{author}{\bibfnamefont{P.}~\bibnamefont{Alonso-González}},
  \bibinfo{author}{\bibfnamefont{S.}~\bibnamefont{Thongrattanasiri}},
  \bibinfo{author}{\bibfnamefont{F.}~\bibnamefont{Huth}},
  \bibinfo{author}{\bibfnamefont{J.}~\bibnamefont{Osmond}},
  \bibinfo{author}{\bibfnamefont{M.}~\bibnamefont{Spasenović}},
  \bibinfo{author}{\bibfnamefont{A.}~\bibnamefont{Centeno}},
  \bibinfo{author}{\bibfnamefont{A.}~\bibnamefont{Pesquera}},
  \bibinfo{author}{\bibfnamefont{P.}~\bibnamefont{Godignon}},
  \bibnamefont{et~al.}, \bibinfo{journal}{Nature}
  \textbf{\bibinfo{volume}{487}}, \bibinfo{pages}{77} (\bibinfo{year}{2012}).

\bibitem[{\citenamefont{Fang et~al.}(2013)\citenamefont{Fang, Thongrattanasiri,
  Schlather, Liu, Ma, Wang, Ajayan, Nordlander, Halas, and Garc{\'\i}a~de
  Abajo}}]{Fang13}
\bibinfo{author}{\bibfnamefont{Z.}~\bibnamefont{Fang}},
  \bibinfo{author}{\bibfnamefont{S.}~\bibnamefont{Thongrattanasiri}},
  \bibinfo{author}{\bibfnamefont{A.}~\bibnamefont{Schlather}},
  \bibinfo{author}{\bibfnamefont{Z.}~\bibnamefont{Liu}},
  \bibinfo{author}{\bibfnamefont{L.}~\bibnamefont{Ma}},
  \bibinfo{author}{\bibfnamefont{Y.}~\bibnamefont{Wang}},
  \bibinfo{author}{\bibfnamefont{P.~M.} \bibnamefont{Ajayan}},
  \bibinfo{author}{\bibfnamefont{P.}~\bibnamefont{Nordlander}},
  \bibinfo{author}{\bibfnamefont{N.~J.} \bibnamefont{Halas}}, \bibnamefont{and}
  \bibinfo{author}{\bibfnamefont{F.~J.} \bibnamefont{Garc{\'\i}a~de Abajo}},
  \bibinfo{journal}{ACS Nano} \textbf{\bibinfo{volume}{7}},
  \bibinfo{pages}{2388} (\bibinfo{year}{2013}).

\bibitem[{\citenamefont{Principi et~al.}(2013)\citenamefont{Principi, Vignale,
  Carrega, and Polini}}]{Principi13}
\bibinfo{author}{\bibfnamefont{A.}~\bibnamefont{Principi}},
  \bibinfo{author}{\bibfnamefont{G.}~\bibnamefont{Vignale}},
  \bibinfo{author}{\bibfnamefont{M.}~\bibnamefont{Carrega}}, \bibnamefont{and}
  \bibinfo{author}{\bibfnamefont{M.}~\bibnamefont{Polini}},
  \bibinfo{journal}{arXiv:1305.4666}  (\bibinfo{year}{2013}).

\bibitem[{\citenamefont{Falkovsky}(2008)}]{falkovsky_optical_2008}
\bibinfo{author}{\bibfnamefont{L.~A.} \bibnamefont{Falkovsky}},
  \bibinfo{journal}{J. Phys.: Conf. Ser.} \textbf{\bibinfo{volume}{129}},
  \bibinfo{pages}{012004} (\bibinfo{year}{2008}).

\bibitem[{\citenamefont{Yan et~al.}(2013)\citenamefont{Yan, Low, Zhu, Wu,
  Freitag, Li, Guinea, Avouris, and Xia}}]{Yan2013}
\bibinfo{author}{\bibfnamefont{H.}~\bibnamefont{Yan}},
  \bibinfo{author}{\bibfnamefont{T.}~\bibnamefont{Low}},
  \bibinfo{author}{\bibfnamefont{W.}~\bibnamefont{Zhu}},
  \bibinfo{author}{\bibfnamefont{Y.}~\bibnamefont{Wu}},
  \bibinfo{author}{\bibfnamefont{M.}~\bibnamefont{Freitag}},
  \bibinfo{author}{\bibfnamefont{X.}~\bibnamefont{Li}},
  \bibinfo{author}{\bibfnamefont{F.}~\bibnamefont{Guinea}},
  \bibinfo{author}{\bibfnamefont{P.}~\bibnamefont{Avouris}}, \bibnamefont{and}
  \bibinfo{author}{\bibfnamefont{F.}~\bibnamefont{Xia}}, \bibinfo{journal}{Nat.
  Phot.} \textbf{\bibinfo{volume}{7}}, \bibinfo{pages}{394}
  (\bibinfo{year}{2013}).

\bibitem[{\citenamefont{Novoselov et~al.}(2005)\citenamefont{Novoselov, Geim,
  Morozov, Jiang, Katsnelson, Grigorieva, Dubonos, and Firsov}}]{Novoselov05}
\bibinfo{author}{\bibfnamefont{K.~S.} \bibnamefont{Novoselov}},
  \bibinfo{author}{\bibfnamefont{A.~K.} \bibnamefont{Geim}},
  \bibinfo{author}{\bibfnamefont{S.~V.} \bibnamefont{Morozov}},
  \bibinfo{author}{\bibfnamefont{D.}~\bibnamefont{Jiang}},
  \bibinfo{author}{\bibfnamefont{M.~I.} \bibnamefont{Katsnelson}},
  \bibinfo{author}{\bibfnamefont{I.~V.} \bibnamefont{Grigorieva}},
  \bibinfo{author}{\bibfnamefont{S.~V.} \bibnamefont{Dubonos}},
  \bibnamefont{and} \bibinfo{author}{\bibfnamefont{A.~A.}
  \bibnamefont{Firsov}}, \bibinfo{journal}{Nature}
  \textbf{\bibinfo{volume}{438}}, \bibinfo{pages}{197} (\bibinfo{year}{2005}).

\bibitem[{\citenamefont{Barnes et~al.}(2003)\citenamefont{Barnes, Dereux, and
  Ebbesen}}]{Barnes03}
\bibinfo{author}{\bibfnamefont{W.~L.} \bibnamefont{Barnes}},
  \bibinfo{author}{\bibfnamefont{A.}~\bibnamefont{Dereux}}, \bibnamefont{and}
  \bibinfo{author}{\bibfnamefont{T.~W.} \bibnamefont{Ebbesen}},
  \bibinfo{journal}{Nature} \textbf{\bibinfo{volume}{424}},
  \bibinfo{pages}{824} (\bibinfo{year}{2003}).

\bibitem[{\citenamefont{Mikhailov}(2011)}]{mikhailov2011}
\bibinfo{author}{\bibfnamefont{S.~A.} \bibnamefont{Mikhailov}},
  \bibinfo{journal}{Phys. Rev. B} \textbf{\bibinfo{volume}{84}},
  \bibinfo{pages}{045432} (\bibinfo{year}{2011}).

\bibitem[{sup()}]{supp}
\bibinfo{note}{See Supplemental Material for more detailed discussions of the
  nonlinear susceptibility, quantization, and nanoribbon-cavity coupling.}

\bibitem[{\citenamefont{Mikhailov and Ziegler}(2008)}]{mikhailov2008}
\bibinfo{author}{\bibfnamefont{S.~A.} \bibnamefont{Mikhailov}}
  \bibnamefont{and} \bibinfo{author}{\bibfnamefont{K.}~\bibnamefont{Ziegler}},
  \bibinfo{journal}{J. Phys.: Condens. Matter} \textbf{\bibinfo{volume}{20}},
  \bibinfo{pages}{384204} (\bibinfo{year}{2008}).

\bibitem[{\citenamefont{Denardo and Putterman}(1988)}]{Denardo88}
\bibinfo{author}{\bibfnamefont{B.}~\bibnamefont{Denardo}} \bibnamefont{and}
  \bibinfo{author}{\bibfnamefont{S.}~\bibnamefont{Putterman}},
  \bibinfo{journal}{Phys. Rev. B} \textbf{\bibinfo{volume}{37}},
  \bibinfo{pages}{3720} (\bibinfo{year}{1988}).

\bibitem[{\citenamefont{Gervasoni and Arista}(2003)}]{Gervasoni03}
\bibinfo{author}{\bibfnamefont{J.~L.} \bibnamefont{Gervasoni}}
  \bibnamefont{and} \bibinfo{author}{\bibfnamefont{N.~R.}
  \bibnamefont{Arista}}, \bibinfo{journal}{Phys. Rev. B}
  \textbf{\bibinfo{volume}{68}}, \bibinfo{pages}{235302}
  (\bibinfo{year}{2003}).

\bibitem[{\citenamefont{Thongrattanasiri
  et~al.}(2012)\citenamefont{Thongrattanasiri, Manjavacas, and Garc{\'\i}a~de
  Abajo}}]{Javier12}
\bibinfo{author}{\bibfnamefont{S.}~\bibnamefont{Thongrattanasiri}},
  \bibinfo{author}{\bibfnamefont{A.}~\bibnamefont{Manjavacas}},
  \bibnamefont{and} \bibinfo{author}{\bibfnamefont{F.~J.}
  \bibnamefont{Garc{\'\i}a~de Abajo}}, \bibinfo{journal}{ACS Nano}
  \textbf{\bibinfo{volume}{6}}, \bibinfo{pages}{1766} (\bibinfo{year}{2012}).

\bibitem[{\citenamefont{Bao and Loh}(2012)}]{Bao12}
\bibinfo{author}{\bibfnamefont{Q.}~\bibnamefont{Bao}} \bibnamefont{and}
  \bibinfo{author}{\bibfnamefont{K.~P.} \bibnamefont{Loh}},
  \bibinfo{journal}{ACS Nano} \textbf{\bibinfo{volume}{6}},
  \bibinfo{pages}{3677} (\bibinfo{year}{2012}).

\bibitem[{\citenamefont{Ruppert et~al.}(2010)\citenamefont{Ruppert, Neumann,
  Kinzel, Krenner, Wixforth, and Betz}}]{Ruppert10}
\bibinfo{author}{\bibfnamefont{C.}~\bibnamefont{Ruppert}},
  \bibinfo{author}{\bibfnamefont{J.}~\bibnamefont{Neumann}},
  \bibinfo{author}{\bibfnamefont{J.~B.} \bibnamefont{Kinzel}},
  \bibinfo{author}{\bibfnamefont{H.~J.} \bibnamefont{Krenner}},
  \bibinfo{author}{\bibfnamefont{A.}~\bibnamefont{Wixforth}}, \bibnamefont{and}
  \bibinfo{author}{\bibfnamefont{M.}~\bibnamefont{Betz}},
  \bibinfo{journal}{Phys. Rev. B} \textbf{\bibinfo{volume}{82}},
  \bibinfo{pages}{081416} (\bibinfo{year}{2010}).

\bibitem[{\citenamefont{Zhan et~al.}(2012)\citenamefont{Zhan, Zhao, Hu, Liu,
  and Zi}}]{Zhan12}
\bibinfo{author}{\bibfnamefont{T.~R.} \bibnamefont{Zhan}},
  \bibinfo{author}{\bibfnamefont{F.~Y.} \bibnamefont{Zhao}},
  \bibinfo{author}{\bibfnamefont{X.~H.} \bibnamefont{Hu}},
  \bibinfo{author}{\bibfnamefont{X.~H.} \bibnamefont{Liu}}, \bibnamefont{and}
  \bibinfo{author}{\bibfnamefont{J.}~\bibnamefont{Zi}}, \bibinfo{journal}{Phys.
  Rev. B} \textbf{\bibinfo{volume}{86}}, \bibinfo{pages}{165416}
  (\bibinfo{year}{2012}).

\bibitem[{\citenamefont{Bludov et~al.}(2012)\citenamefont{Bludov, Peres, and
  Vasilevskiy}}]{Bludov12}
\bibinfo{author}{\bibfnamefont{Y.~V.} \bibnamefont{Bludov}},
  \bibinfo{author}{\bibfnamefont{N.~M.~R.} \bibnamefont{Peres}},
  \bibnamefont{and} \bibinfo{author}{\bibfnamefont{M.~I.}
  \bibnamefont{Vasilevskiy}}, \bibinfo{journal}{Phys. Rev. B}
  \textbf{\bibinfo{volume}{85}}, \bibinfo{pages}{245409}
  (\bibinfo{year}{2012}).

\bibitem[{\citenamefont{Ferreira and Peres}(2012)}]{Aires12}
\bibinfo{author}{\bibfnamefont{A.}~\bibnamefont{Ferreira}} \bibnamefont{and}
  \bibinfo{author}{\bibfnamefont{N.~M.~R.} \bibnamefont{Peres}},
  \bibinfo{journal}{Phys. Rev. B} \textbf{\bibinfo{volume}{86}},
  \bibinfo{pages}{205401} (\bibinfo{year}{2012}).

\bibitem[{\citenamefont{Nikitin
  et~al.}(2012{\natexlab{b}})\citenamefont{Nikitin, Guinea, and
  Mart{\'\i}n-Moreno}}]{Nikitin12b}
\bibinfo{author}{\bibfnamefont{A.~Y.} \bibnamefont{Nikitin}},
  \bibinfo{author}{\bibfnamefont{F.}~\bibnamefont{Guinea}}, \bibnamefont{and}
  \bibinfo{author}{\bibfnamefont{L.}~\bibnamefont{Mart{\'\i}n-Moreno}},
  \bibinfo{journal}{Appl. Phys. Lett.} \textbf{\bibinfo{volume}{101}},
  \bibinfo{pages}{151119} (\bibinfo{year}{2012}{\natexlab{b}}).

\bibitem[{\citenamefont{Zhu et~al.}(2013)\citenamefont{Zhu, Yan, Uhd~Jepsen,
  Hansen, Asger~Mortensen, and Xiao}}]{Zhu13}
\bibinfo{author}{\bibfnamefont{X.}~\bibnamefont{Zhu}},
  \bibinfo{author}{\bibfnamefont{W.}~\bibnamefont{Yan}},
  \bibinfo{author}{\bibfnamefont{P.}~\bibnamefont{Uhd~Jepsen}},
  \bibinfo{author}{\bibfnamefont{O.}~\bibnamefont{Hansen}},
  \bibinfo{author}{\bibfnamefont{N.}~\bibnamefont{Asger~Mortensen}},
  \bibnamefont{and} \bibinfo{author}{\bibfnamefont{S.}~\bibnamefont{Xiao}},
  \bibinfo{journal}{Appl. Phys. Lett.} \textbf{\bibinfo{volume}{102}},
  \bibinfo{pages}{131101} (\bibinfo{year}{2013}).

\bibitem[{\citenamefont{Snyder and Love}(1983)}]{Snyder83}
\bibinfo{author}{\bibfnamefont{A.~W.} \bibnamefont{Snyder}} \bibnamefont{and}
  \bibinfo{author}{\bibfnamefont{J.}~\bibnamefont{Love}},
  \emph{\bibinfo{title}{Optical Waveguide Theory}}
  (\bibinfo{publisher}{Springer}, \bibinfo{year}{1983}), \bibinfo{edition}{1st}
  ed.

\bibitem[{\citenamefont{Liao and Law}(2010)}]{Liao10}
\bibinfo{author}{\bibfnamefont{J.-Q.} \bibnamefont{Liao}} \bibnamefont{and}
  \bibinfo{author}{\bibfnamefont{C.~K.} \bibnamefont{Law}},
  \bibinfo{journal}{Phys. Rev. A} \textbf{\bibinfo{volume}{82}},
  \bibinfo{pages}{053836} (\bibinfo{year}{2010}).

\bibitem[{\citenamefont{Eisaman et~al.}(2011)\citenamefont{Eisaman, Fan,
  Migdall, and Polyakov}}]{Eisaman11}
\bibinfo{author}{\bibfnamefont{M.~D.} \bibnamefont{Eisaman}},
  \bibinfo{author}{\bibfnamefont{J.}~\bibnamefont{Fan}},
  \bibinfo{author}{\bibfnamefont{A.}~\bibnamefont{Migdall}}, \bibnamefont{and}
  \bibinfo{author}{\bibfnamefont{S.~V.} \bibnamefont{Polyakov}},
  \bibinfo{journal}{Rev. Sci. Inst.} \textbf{\bibinfo{volume}{82}},
  \bibinfo{pages}{071101} (\bibinfo{year}{2011}).

\bibitem[{\citenamefont{Marsili et~al.}(2012)\citenamefont{Marsili, Bellei,
  Najafi, Dane, Dauler, Molnar, and Berggren}}]{Marsili12}
\bibinfo{author}{\bibfnamefont{F.}~\bibnamefont{Marsili}},
  \bibinfo{author}{\bibfnamefont{F.}~\bibnamefont{Bellei}},
  \bibinfo{author}{\bibfnamefont{F.}~\bibnamefont{Najafi}},
  \bibinfo{author}{\bibfnamefont{A.~E.} \bibnamefont{Dane}},
  \bibinfo{author}{\bibfnamefont{E.~A.} \bibnamefont{Dauler}},
  \bibinfo{author}{\bibfnamefont{R.~J.} \bibnamefont{Molnar}},
  \bibnamefont{and} \bibinfo{author}{\bibfnamefont{K.~K.}
  \bibnamefont{Berggren}}, \bibinfo{journal}{Nano Lett.}
  \textbf{\bibinfo{volume}{12}}, \bibinfo{pages}{4799} (\bibinfo{year}{2012}).

\bibitem[{\citenamefont{Karasik et~al.}(2012)\citenamefont{Karasik, Pereverzev,
  Soibel, Santavicca, Prober, Olaya, and Gershenson}}]{Karasik12}
\bibinfo{author}{\bibfnamefont{B.~S.} \bibnamefont{Karasik}},
  \bibinfo{author}{\bibfnamefont{S.~V.} \bibnamefont{Pereverzev}},
  \bibinfo{author}{\bibfnamefont{A.}~\bibnamefont{Soibel}},
  \bibinfo{author}{\bibfnamefont{D.~F.} \bibnamefont{Santavicca}},
  \bibinfo{author}{\bibfnamefont{D.~E.} \bibnamefont{Prober}},
  \bibinfo{author}{\bibfnamefont{D.}~\bibnamefont{Olaya}}, \bibnamefont{and}
  \bibinfo{author}{\bibfnamefont{M.~E.} \bibnamefont{Gershenson}},
  \bibinfo{journal}{Appl. Phys. Lett.} \textbf{\bibinfo{volume}{101}},
  \bibinfo{pages}{052601} (\bibinfo{year}{2012}).

\bibitem[{\citenamefont{Fetter}(1973)}]{Fetter73}
\bibinfo{author}{\bibfnamefont{A.}~\bibnamefont{Fetter}},
  \bibinfo{journal}{Ann. Phys. N.Y.} \textbf{\bibinfo{volume}{81}},
  \bibinfo{pages}{367} (\bibinfo{year}{1973}).

\end{thebibliography}

\setcounter{equation}{0}
\renewcommand{\theequation}{S.\arabic{equation}}
\setcounter{figure}{0}
\renewcommand{\thefigure}{S.\arabic{figure}}

\appendix
\newpage
\section{Supplemental Material}
\emph{Nonlinear conductivity} -
 The nonlinearity can be derived from the Boltzmann equation for the 2D electron distribution function $f(\bm{x},\bm{p},t)$, which is a function of two space $\bm{x}=(x,y)$ and two momentum $\bm{p}=(p_x,p_y)$ variables and time $t$,  and Poisson's equation for the electric potential $\varphi(\bm{x},z,t)$
\begin{align}
\del_t f +v_F\hat{p} \cdot &\del_{\bm{x}} f+ e \del_{\bm{x}} \varphi \cdot \del_{\bm{p}} f = 0 \\
(\del_{\bm{x}}^2+\del_z^2) \varphi &= e\, n \, \delta(z) /\epsilon_0 
\end{align}
where the 2D electron density is defined as  $n \equiv \int d \bm{k} f$, $\del_{\bm{x}}=\del_x \hat{x}+\del_y \hat{y}$, and $\del_{\bm{p}}= \del_{p_x}\hat{x}+\del_{p_y}\hat{y}$.  Eq.~S1 is written for electron doping, for hole doping the equations and final results would be the same after taking the opposite electron charge.  These equations are an extension of the Drude model to describe intraband transitions. Taking $x$ to be the propagation direction, the nonlinear equations for the electron density $n$ and current density $n \,\bm{\bar{v}}= \int d \bm{k}\, v_F \hat{k} \, f $ can be derived as
\begin{align} \label{eqn:nDot}
\del_t n& + \del_x n\, \bar{v} = 0 \\ \label{eqn:vbarDot}
\del_t \bar{v}&  - \frac{e}{m^*} \del_x \varphi + \bar{v}\, \del_x \bar{v} +\frac{3 e}{2 m^*n_0} \del_x \varphi\, \delta n = 0
\end{align}
where the effective mass for the plasmon excitations is $m^*=\hbar k_F/v_F$, $\delta n= n-n_0$, and $n_0 \equiv k_F^2/\pi$ is the equilibrium electron density.
Linearizing these equations around $n_0$ and $\bar{v}=0$ gives the plasmon dispersion from Eq.~1 in the main text \cite{Fetter73}.  The nonlinearity is described by the last two terms in Eq.~S\ref{eqn:vbarDot}, where the second term, $\propto\del \varphi \delta n$, arises from the linear band structure in graphene and is absent for electrons with a parabolic dispersion.  To find the nonlinear conductivity, we can expand $\bar{v}$ and $n$ in spatial Fourier components and solve the resulting coupled equations together with Poisson's equation \cite{Denardo88}.  This allows us to express $\sigma^{3}(\omega)$ through the identity $e n \bar{v} = \sigma(\omega) E + \sigma^{3}(\omega) E^3$.

To solve for the nonlinear shift in the cavity, we use the boundary condition that the current density perpendicular to the edge is zero at the edge.  This allows us to represent $\bar{\bm{v}} = \hat{x}\sum_p v_p \sin p x$ and $n= \sum_q n_q \cos q x$ where $q=m k_{sp}$ for some integer $m$.
Inserting this solution into Eq.~S\ref{eqn:nDot}-\ref{eqn:vbarDot} leads to coupled nonlinear equations for $n_q$ and $v_q$
\begin{align}  \nonumber
\sum_p &\sin p x \bigg(\dot{v}_p - \frac{\omega_p^2}{n_0 p} n_p \bigg) = \frac{1}{2} \sum_{p,q}[ p v_p v_{q} \sin (p-q) x \\
-&\bigg( p v_p v_{q} - \frac{3}{2}\frac{\omega_p^2}{p} \frac{n_p n_{q}}{n_0^2}\bigg) \sin (p+q) x ]\\\nonumber
\sum_p& \cos px ( \dot{n}_p +n_0 p v_p ) =  \frac{1}{2} \sum_{p,q} n_p v_q \big[ (p-q) \cos (p-q)x
\\
 -& (p+q) \cos (p+q)x \big]
\end{align}
where $\omega_p^2= \frac{e^2\omega_F}{2 \pi \epsilon_0  \hbar^2}   p$ .  These equations can be solved in perturbation theory to find the nonlinear frequency shift of the plasmon resonance as
\be
\delta \omega_p= \frac{\omega_p}{16} \frac{\delta n^2}{n_0^2} \bigg( \frac{5 \omega_p^2+ \omega_{2p}^2}{4 \omega_p^2 - \omega_{2p}^2} - \frac{2 \omega_p^2}{\omega_{2 p}^2} \bigg) \ge \frac{ 5}{32} \omega_p \frac{\delta n^2}{n_0^2}
\ee

\emph{Quantizing the plasmon mode} -
To quantize the plasmon mode we use the Hamiltonian \cite{Gervasoni03}
\be
\begin{split}
H &= \frac{1}{2} \int d \bm{x}\, e\,\delta  n\, \varphi + \frac{1}{2} \int d\bm{x} n_0 m^* {\bar{v}}^2 \\
 &=\frac{A\, m^*}{4\, n_0} \sum_q \frac{1}{q^2} (\omega_q^2\, \delta n_q^2+ \delta \dot{n}_q^2)
\end{split}
\ee
where $A=\pi^2/k_{sp}^2$ is the area of the graphene sheet and we used the relation $\bar{v}_q =- \delta \dot{n}_q/q\, n_0$ from the continuity equation.  This Hamiltonian can be quantized in the usual way by defining $\delta n_q = \frac{\gamma_q}{2 \omega_q} (a_q+ a_q^\dagger)$ for bosonic operators $a_q$ such that $\dot{a}_q=- i \omega_q a_q$ and $\gamma_q = 2 q \sqrt{\omega_q \omega_F/\pi A}$.  This leads directly to Eq.~4 in the main text.

\emph{Coupling between nanoribbon and cavity} -
To calculate the coupling between the cavity and the proximal nanoribbion we use the electric potential of the nanoribbon plasmons acting on the graphene cavity
\be
\begin{split}
\varphi_r(x)& =\frac{1}{4 \pi \epsilon_0 }\sum_k \int d\bm{x}' \frac{e\, n^r_k\, \cos k x'}{\lvert x+d-x' \lvert } \\
&\approx  \frac{W_r}{4 \pi \epsilon_0 }\sum_k \frac{e \, n^r_k}{k^2 (x+d)^2}
\end{split}
\ee
where we assumed $d\gg W, \lambda_{sp}$.
Inserting this into Eq.~S\ref{eqn:vbarDot} gives the coupling between each plasmon mode $k$ in the nanoribbon with the plasmon mode $q$ of the cavity as
\be
\kappa_{kq}= \frac{8}{\pi} \sqrt{ \frac{k_F^r \, W_r}{k_F^c L} } \frac{\omega_q^{c\,2}}{\omega^r_k+\omega^c_q} \frac{1}{q^2 k\, d^3}
\ee
where $L$ is the length of the nanoribbon, $k_F^{r,c}$ is the Fermi wavevector and $\omega_k^{r,c}$ is the dispersion of the ribbon($r$) and cavity($c$).  Applying Fermi's Golden rule gives the decay rate of the cavity mode into the nanoribbon plasmons given in Eq.~8.

\end{document}